\documentclass[journal]{IEEEtran}
\usepackage{algorithm}
\usepackage{algorithmic}
\usepackage{amsmath,epsfig}
\usepackage{amssymb,float}
\usepackage{cite}
\usepackage{color}
\usepackage{bm}
\usepackage{multirow}
\newcommand{\thickhline}{%
    \noalign {\ifnum 0=`}\fi \hrule height 1pt
    \futurelet \reserved@a \@xhline}
\IEEEoverridecommandlockouts
\begin{document}
\title{Algorithmic Obfuscation for LDPC Decoders}
\author{\IEEEauthorblockN{Jingbo Zhou and Xinmiao Zhang,~\IEEEmembership{Senior Member,~IEEE}}\\

\thanks{
This work is supported in part by US Air Force Research Laboratory under Award Number: FA8650-20-C-1719

The authors are with The Ohio State University, Columbus, OH 43210, USA. Emails: \{zhou.2955, zhang.8952\}@osu.edu.}}

\IEEEtitleabstractindextext{%
\begin{abstract}
In order to protect intellectual properties against untrusted foundry, many logic-locking schemes have been developed. {The idea of logic locking is to insert a key-controlled block into the circuit to make the circuit function incorrectly or go through redundant states without right keys. However, in the case that the algorithm implemented by the circuit is self-correcting, existing logic-locking schemes do not affect the system performance much even if a wrong key is used and hence do not effectively protect the circuit.} One example is low-density parity-check (LDPC) error-correcting decoders, which are used in numerous digital communication and storage systems. This paper proposes two algorithmic-level obfuscation methods for LDPC decoders. By modifying the decoding process and locking the stopping criterion, our new designs substantially degrade the decoder throughput and/or error-correcting performance, and make the decoder unusable when a wrong key is applied. For an example LDPC decoder, our proposed methods reduce the throughput to less than 1/3 and/or increase the decoder error rate by at least two orders of magnitude with at most 0.55\% area overhead. Besides, our designs are also resistant to the SAT, AppSAT and removal attacks.
\end{abstract}

}

\maketitle

\IEEEdisplaynontitleabstractindextext
\IEEEpeerreviewmaketitle

\section{Introduction}

\IEEEPARstart{L}{ogic} locking \cite{RoyEPIC} helps to protect intellectual properties (IPs) of circuits by inserting key-controlled components. The circuit does not function correctly without the right key. {Among oracle-guided attacks, the satisfiability-based (SAT) attacks \cite{SAT} are the most powerful \cite{approximation}.} The main idea of the SAT attack is to apply Boolean SAT solvers iteratively to find distinguishing input patterns (DIPs), which are used to exclude wrong keys. Many logic-locking schemes, such as those in \cite{RoyEPIC, RajendranSecurity, AND logic, Fault analysis, Lee-logic, LUT2} have been decrpted by the SAT attack. The Anti-SAT \cite{Anti-SAT} and SARlock \cite{SARlock} schemes make the number of iterations needed by the SAT attack exponential. However, both of them are composed of large AND/NAND trees, which make them subject to the approximate (App-) SAT attack \cite{AppSAT}. In addition, the large AND/NAND trees can be identified and replaced by removal attacks \cite{removal,ISCAS removal}. By increasing the corruptibility of all wrong keys, the Diversified Tree Logic (DTL) design \cite{AppSAT} achieves better resistance to the AppSAT attack at the cost of degraded resilience to the SAT attack. Different from the Anti-SAT and SARlock designs that have no overlap among the sets of wrong keys excluded by different DIPs, the Generalized (G-) Anti-SAT design \cite{G-Anti-SAT} allows overlap among the wrong key sets. It can be built by using  a large number of possible functions instead of AND/NAND trees and is more resilient to removal attacks. Besides, it achieves better resistance to the AppSAT attack without sacrificing the resistance to the SAT attack. The recently proposed Cascade (CAS)-Lock scheme \cite{CAS} is a special case of the G-Anti-SAT design.

The stripped functionality logic locking (SFLL) \cite{SFLL} scheme corrupts the original circuit and adds a logic-locking block to correct the corrupted circuit when the right key is applied. The circuit does not function correctly if the key-controlled block is removed and its output is replaced by a constant signal. Hence, the SFLL can more effectively resist removal attacks. However, the functional analysis on logic locking (FALL) \cite{FALL} and Hamming distance (HD)-unlock method \cite{HD-unlock} can successfully attack the SFLL by analyzing its functional and structural properties. 

In systems that are fault-tolerating or self-correcting, errors forced on the signals may not affect the overall output or may get corrected by later computations. Examples are low-density parity-check (LDPC) codes \cite{first LDPC}, machine learning, and applications that are amiable to approximate computing. For the circuits implementing these algorithms, even if a logic-locking scheme is used and a signal is flipped when wrong keys are applied, the errors resulted at the overall output may be negligible. To substantially increase the output error when the key is incorrect, the obfuscation scheme needs to be designed using the properties of the algorithm. For circuits implementing machine learning, algorithmic logic locking has been developed in \cite{ML} by utilizing the statistics of the inputs. Besides, obfuscation keys are used in \cite{DNNProtect} to swap the filters or rows/columns of the filters in convolutional neural networks. Since these techniques depend on the specifics of neural networks, they do not apply to other algorithms.

{In this paper, for the first time, algorithmic approaches are investigated to obfuscate LDPC decoders, which are self-correcting. LDPC codes are among the most extensively-used error-correcting codes in digital communications and storage. They are utilized in most of the controllers for Flash memories and magnetic drives. They can be also found in numerous communication applications such as 5G/6G wireless communications, Wi-Fi, digital video broadcasting, and Ethernet. LDPC codes can be decoded by different algorithms with tradeoffs on error-correcting performance and complexity. Even for the same decoding algorithm, various architectures and scheduling schemes can be used to achieve different throughput and area complexity \cite{MyBook}. Additionally, a variety of techniques, such as dynamic message scaling, decoding re-trial, and post-processing, can be utilized to handle the error-floor problem of LDPC codes, which is a harmful phenomena that the error-correcting performance curve flattens as the input error rate decreases \cite{LinBook}. As a result, LDPC decoder architectures optimized for target applications consist of many proprietary designs and the corresponding development is very time-consuming. Hence, it is very important to protect the IP of LDPC decoders.} 

Given the probability information from the communication or storage channel, in general, LDPC decoder iteratively updates probability messages according to the parity check matrix that defines the code. At the end of each iteration, the decoder checks if a codeword is found to decide whether to stop. Even if message bits get flipped, later decoding iterations may not be affected at all.

This paper proposes two algorithmic-level obfuscation methods for LDPC decoders. Both of them replace the original stop condition checking part with a key controlled logic-locking block and effectively obfuscate the decoder as follows:
\begin{enumerate}
    \item {The first proposed scheme makes the decoding process run until the last iteration in most cases when a wrong key is used. Accordingly, the decoder throughput is reduced by several times. If the decoder can not meet the system throughput requirement, then it becomes not usable. }
    
    \item {The second proposed scheme uses more complicated logic compared to that of the first design to make most of the wrong keys have high corruptibility. As a result, it leads to substantial degradation on not only the throughput but also the error-correcting performance when the key is incorrect.}
\end{enumerate}
Additionally, low-overhead modifications on the LDPC decoding algorithm are developed in this paper to change the correct decoding stop condition so that the right key can not be guessed. The proposed schemes effectively resist the SAT and AppSAT attacks. Even if an approximate key is returned by the attacks, it makes the decoder have several times lower throughput and become not usable. Our design is also resistant to removal attacks. Despite that the logic-locking block may be identified from the netlist, the attacker is not able to replace it to make the decoder function correct since the right stop condition has been modified and is unknown to the attacker. {Take the decoder of a (1270, 635) LDPC code that encodes 635-bit data into 1270-bit codewords as an example}, our proposed methods reduce the throughput to less than 1/3 and/or increase the decoder error rate by at least two orders of magnitude at practical channel conditions. The proposed designs only bring 0.15\% and 0.55\%, respectively, area overheads to the LDPC decoder without any penalty on the achievable clock frequency. 


This paper is organized as follows. Section II briefly introduces major attacks, existing logic-locking schemes, and LDPC codes. The two proposed algorithmic obfuscation methods for LDPC decoders are detailed in Section III. The security and complexity overhead analyses of the proposed schemes are presented in Section IV. Discussions and conclusions follow in Section V and VI, respectively.

\section{Backgrounds and Motivations}
This section first introduces the attack model, basic knowledge on logic locking, various attacks, and LDPC codes. Then it is highlighted that existing logic-locking schemes cannot effectively degrade the performance of LDPC decoders even if a wrong key is used. 

\subsection{Attack model}
The attack model assumes that the attacker has access to: 1) the netlist of the locked chip. Besides, signal probability skew computation and other analyses can be carried out using the netlist; 2) a functioning chip with the correct key input. However, the attacker does not have perfect knowledge on every detail or the dataflow of the algorithm implemented in the chip. These are the same assumptions made in previous papers \cite{SAT,Anti-SAT,SARlock,AppSAT,removal,G-Anti-SAT}.

\subsection{Logic locking and attacks}
\begin{figure}
    \centering
    \includegraphics[width=2.0in]{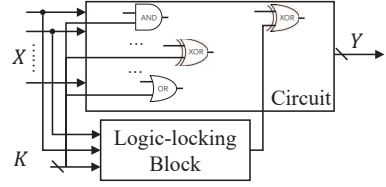}
    \caption{Obfuscated circuit using existing logic-locking schemes}
    \label{LogicLock}
\end{figure}

{Fig. \ref{LogicLock} shows the idea of existing logic-locking schemes. Earlier schemes directly use key inputs to flip signals through AND, OR, and XOR gates \cite{RoyEPIC, RajendranSecurity, AND logic, Fault analysis, Lee-logic, LUT2}. These designs are subject to the SAT attack \cite{Anti-SAT}. To resist the SAT attack, a carefully designed key-controlled logic-locking block can be inserted as shown in Fig. \ref{LogicLock}. Its output is XORed with a signal that affects the overall output most. The output of the circuit is corrupted under certain input patterns when a wrong key is used. Only when the right key handed out to the authorized user is applied, the circuit functions correctly.}

{The SAT attack is a major threat to logic locking. The attacker is assumed to have a functioning chip with the correct key inserted and the netlist of the locked circuit} represented by $Y = f(X, K)$, where $X$, $K$, and $Y$ are the primary input, key input, and primary output vector, respectively. The circuit can be described by a conjunctive normal form (CNF) formula. The SAT attack is an iterative process. In each iteration, it uses the SAT solver to find a distinguishing input pattern (DIP), which is an input vector that leads to different outputs under different keys. The DIP is utilized to query the functioning chip to get the corresponding original output. Then the CNF formula is updated accordingly so that the keys making the circuit output different from the original output are excluded. When no more DIP can be found, the SAT attack stops. At this point, all the wrong keys have been excluded, and the right keys are derived.


{The Anti-SAT and SARLock logic-locking schemes in \cite{Anti-SAT, SARlock} were proposed to resist the SAT attack. By making each input pattern exclude unique wrong keys, the number of iterations needed by the SAT attack is exponential to the number of bits in data input. However, this also means that any random key only leads to the wrong output for one single input pattern.}

The AppSAT attack \cite{AppSAT} is developed based on the SAT attack. In this attack, after every certain number of iterations, random input patterns are utilized to query the functioning chip. The portion of the input patterns generating the wrong output is calculated and the constraints from these queries are also added to the CNF formula each time. If the portion falls below a threshold for a number of rounds, the AppSAT attack terminates and returns a key from the remaining keys as an approximate key. Define the corruptibility of a wrong key as the number of input patterns that make the output different from the original output when the wrong key is applied. The wrong keys with higher corruptibility are more likely to be excluded by the AppSAT attack. If the returned key has very low corruptibility, then it can be used as an approximate key.

{In order to increase the corruptibility of wrong keys, key bits directly connected to intermediate signals in the circuit through AND/OR/XOR gates can be utilized in addition to the logic-locking block as shown in Fig. \ref{LogicLock} \cite{Anti-SAT}. However, these extra key bits do not increase the number of iterations needed by the SAT attack \cite{Anti-SAT}. Besides, these key bits can be recovered by the AppSAT attack after a small number of iterations \cite{AppSAT}. The recent G-Anti-SAT scheme \cite{G-Anti-SAT} makes the corruptibility of a large portion of the wrong keys tunable without sacrificing the resistance to the SAT attack. As a result, even with a large number of iterations in the AppSAT attack, the probability that a low-corruptibility wrong key is returned is small. To effectively resist the AppSAT attack, the DTL design \cite{AppSAT} increases the corruptibility of all wrong keys at the cost of sacrificed SAT-attack resilience.}

{Logic-locking schemes can be also applied to lock the finite state machine (FSM) of a circuit. Examples include HARPOON \cite{HARPOON} and the parametric locking in \cite{YasinPara}. HARPOON modifies the original FSM to corrupt the primary output of the circuit and it is subject to the SAT attack \cite{ORACALL}. The parametric locking inserts dummy states to increase the number of clock cycles spent on the computations. However, the dummy states can be removed by resynthesis attack \cite{YasinPara}.}

\subsection{LDPC codes and decoding algorithms}

\begin{figure}
    \centering
    \includegraphics[width=3.3in]{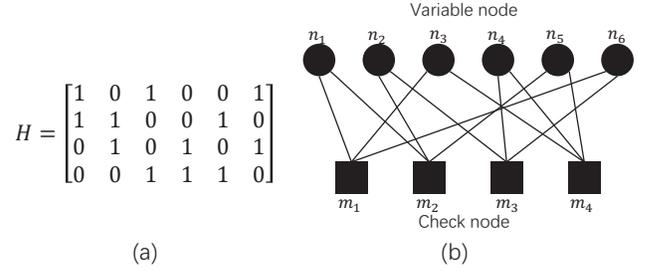}
    \caption{(a). A toy example of parity check matrix; (b) Tanner graph of the example parity check matrix}
    \label{H}
\end{figure}

{LDPC codes are linear block error-correcting codes, which add redundancy to the messages in the encoding process. If the number of errors does not exceed the correction capability of the code, they can be corrected at the receiver using a decoding process.} LDPC codes can be specified by the parity check matrix $H$. An $(N,S)$ code encodes $S$-bit messages to $N$-bit codewords, where $S < N$. {If $H$ is full-rank, its dimension is $(N-S)\times N$. A vector $c = [c_0, c_1, \cdots, c_{N-1}]$ is a codeword iff $cH^T=0$. Here the length of the zero vector is $N-S$ bits.} A toy example of the parity check matrix is shown in Fig. \ref{H}(a). For real LDPC codes used in practical systems, $N$ is at least several thousands and the parity check matrix $H$ is large and very sparse. The $H$ matrix can be also represented by a Tanner graph as shown in Fig. \ref{H}(b). Each row of $H$ is represented by a check node in the Tanner graph and specifies a check equation that needs to be satisfied. Each column of $H$ corresponds to a variable node. A nonzero entry in $H$ represents the connection edge between the corresponding check node and variable node. 

Due to the noise in the communication or storage channel, there are errors in the received bits and such error rate is decided by the application. The inputs to the LDPC decoder are the probabilities that each received bit is `1' from the channel information. During the decoding process, probability messages are passed between the connected variable and check nodes and got updated iteratively.
At the end of each decoding iteration, hard decisions on whether each received bit should be '1' or '0'  are made based on the most updated messages. The decoding stops when the hard-decision vector is a codeword. Otherwise, the decoding process continues until a pre-set maximum iteration number, $I_{max}$, in which case decoding failure is declared. The error-correcting performance of LDPC decoder improves with $I_{max}$. However, the additional gain becomes negligible for large $I_{max}$. On the other hand, using a large $I_{max}$ increases the worst-case decoding latency and reduces the throughput. Hence, $I_{max}$ is typically set to a moderate number, such as 15, to achieve good error-correcting performance and low latency. 

LDPC codes can be decoded by many different decoding algorithms \cite{MyBook}. The most popular one is the Min-sum algorithm \cite{Minsum}, since it can well balance the complexity and error-correcting performance. This algorithm is chosen to illustrate our proposed designs and its pseudo codes are listed in Algorithm \ref{Minsum}. The input to the decoder is the channel information $\gamma_n$, which is the log-likelihood ratio of the probability that the $n$-th input bit is `1' over the probability that it equals `0'. The decoder output is the $z$ vector when the decoding stops. In this algorithm, the message from variable node $n$ to check node $m$ is denoted by $u_{m,n}$. The message from check node $m$ to variable node $n$ is $v_{m,n}$. $S_c(n)$ represents the set of check nodes connected to variable node $n$ and $S_v(m)$ denotes the set of variable nodes connected to check node $m$. {Take the variable node $n_1$ and check node $m_1$ shown in Fig. \ref{H}(b) as examples. $S_c(n_1)$ equal $\{m_1, m_2\}$ and $S_v(m_1)$ is $\{n_1, n_3, n_6\}$.} 

\begin{algorithm}[t]
\caption{The Min-sum Decoding Algorithm}
\label{Minsum}
\begin{algorithmic}[1]
\STATE { {\textbf{input:} $\gamma_n$ {=probability that bit $n$ is `1' from channel}}}
\STATE {\textbf{initialization:} $u_{m,n} = \gamma_n; z_n = sign(\gamma_n)$} 
\STATE { {if $zH^T = 0$}}
\STATE{\quad {return codeword $z$ and stop}}
\STATE{for $i = 1$ to $I_{max}$}
\STATE {\quad \textbf{check node processing}}
\STATE {\quad {for each $m$}}
\STATE {\quad \quad $min1_m = min_{j\in S_v(m)}|u_{m,j}|$}
\STATE {\quad \quad $idx_m = argmin_{j\in S_v(m)}|u_{m,j}|$}
\STATE {\quad \quad $min2_m = min_{j\in S_v(m),\ j\neq idx_m}|u_{m,j}|$}
\STATE {\quad \quad $s_m = \prod_{j\in S_v(m)}sign(u_{m,j})$}
\STATE{\quad \quad for each $n \in S_v(m)$}
\STATE {\quad \quad \hspace{2em}$|v_{m,n}| =  \left\{
             \begin{array}{lr}
             \alpha min1_m\ if\  n \neq idx_m &  \\
             \alpha min2_m\ if\  n = idx_m & \\  
             \end{array}
\right.$}
\STATE {\quad \quad \hspace{2em}$sign(v_{m,n}) = s_m sign(u_{m,n})$}
\STATE{\quad \textbf{variable node processing}}
\STATE {\quad {for each $n$ and $m$}}
\STATE{\quad \quad $u_{m,n} = \gamma_n + \sum_{i\in S_c(n),i\neq m}v_{i,n}$}
\STATE{\quad \textbf{a posteriori info. comp. \& tentative decision}}
\STATE {\quad {for each $n$}}
\STATE{\quad \quad $\Tilde{\gamma_n} = \gamma_n + \sum_{i\in S_c(n)}v_{i,n}$}
\STATE{\quad \quad $z_n = sign(\Tilde{\gamma_n})$}
\STATE {\quad  {if $zH^T = 0$ }}
 \STATE {\quad\quad  {output codeword $z$ and stop}}
\STATE{end for}
\STATE{ {output $z$ and declare decoding failure}}
\end{algorithmic}
\end{algorithm}

{In the beginning of the Min-sum decoding, the variable-to-check (v2c) messages, $u_{m,n}$, are initialized and hard decisions, $z_{n}$, are first made according to the channel input $\gamma_{n}$ in Line 2. For example, $u_{m_1,n_1}=u_{m_2,n_1}=\gamma_1$ and $u_{m_2,n_2}=u_{m_3,n_2}=\gamma_2$ for the toy code shown in Fig. \ref{H}. Then the product of the initial hard-decision vector, $z$, and $H^T$ is calculated. If it is zero, there is no error and $z$ is returned as the codeword. Otherwise, the decoding iteration starts.}

{Each decoding iteration is divided into three steps: check node processing, variable node processing, and {\it a posteriori} information computation. In the processing for check node $m$, the smallest and second smallest magnitudes among all the messages from the connected variable nodes are computed in Lines 8-10 of Algorithm \ref{Minsum}. They are denoted by $min1_m$ and $min2_m$, respectively. Besides, $idx_m$ is the index of the connected variable node whose message magnitude is the smallest. Take the $m_1$ node in Fig. \ref{H}(b) as an example. It is connected to three variable nodes $n_1$, $n_3$, and $n_6$. Hence, $min1_{m_1}$ and $min2_{m_1}$ are the smallest and second smallest, respectively, among $|u_{m_1, n_1}|, |u_{m_1, n_3}|$, and $|u_{m_1, n_6}|$. Additionally, the product of all the signs of the messages from the variable nodes connected to check node $m$ is computed as $s_m$ in Line 11 of Algorithm \ref{Minsum}. Then the check-to-variable (c2v) messages, $v_{m,n}$, are computed according to Lines 12-14 of Algorithm \ref{Minsum}. The magnitude of  $v_{m,n}$ equals either $min1_m$ or $min2_m$ scaled by a factor $\alpha$ depending on whether $n$ equals $idx_m$. The pre-determined scalar $ 0 < \alpha < 1$ is used to improve the error-correcting performance. In the variable node processing step shown in Lines 16-17 of Algorithm 1, the updated v2c message to be sent to check node $m$ in the next iteration, $u_{m, n}$, is calculated as the sum of the channel input for variable node $n$ and the messages from all connected check nodes, except the check node $m$ itself. The {\it a posteriori} information, $\tilde{\gamma_{n}}$ is computed in a similar way as $u_{m,n}$, except that the messages from all connected check nodes are added. Take the variable node $n_3$ in the toy code shown in Fig. \ref{H} as an example. It is connected to two check nodes $m_1$ and $m_4$. Hence $u_{m_1, n_3}=\gamma_{3}+v_{m_4,n_3}$ and $u_{m_4, n_3}=\gamma_{3}+v_{m_1,n_3}$. Besides, $\tilde{\gamma_{3}}=\gamma_{3}+v_{m_1,n_3}+v_{m_4,n_3}$. The hard decision $z$ is made again at the end of each iteration as the signs of $\tilde {\gamma_n}$. If $zH^T = 0$ is satisfied, $z$ is the recovered codeword. The decoding stops and $z$ is sent to the decoder output. Otherwise, the decoding continues to the next iteration. If no codeword is found in $I_{max}$ iterations, decoding failure is declared. }

\begin{figure}
    \centering
    \includegraphics[width=3.5in]{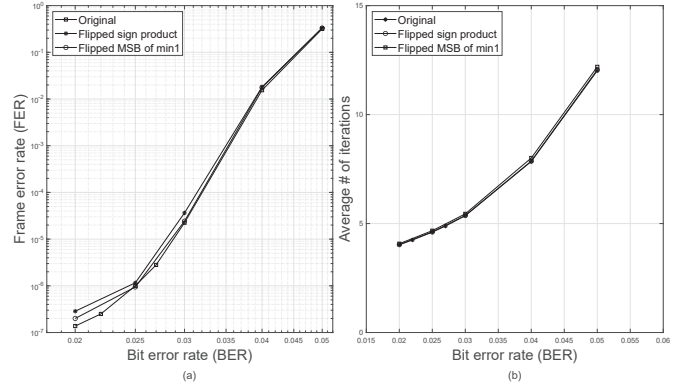}
    \caption{Simulation results for (1270, 635) QC-LDPC code: (a) FERs; (b) average number of iterations for original decoder and decoders with sign product and MSB of $min1$ flipped for one check node in every iteration with $I_{max}=15$}
    \label{LDPC_comp}
\end{figure}

{The frame error rate (FER) and average number of iterations, which decides the throughput, are the two criteria used to measure the performance of the LDPC decoder. If the FER or throughput can not meet the system requirement, then the LDPC decoder is not usable. Quasi-cyclic (QC)-LDPC codes \cite{QC-LDPC} are usually used in practical applications because the structure of their $H$ matrices allows efficient parallel processing. For an example (1270, 635)  QC-LDPC code whose $H$ matrix consists of $5\times 10$ shifted identity matrices of dimension $127\times 127$, simulation results for the FER and average number of decoding iterations are plotted in Fig. \ref{LDPC_comp} for a range of input bit error rates (BERs). In our simulations, binary symmetric channel is assumed and $I_{max}$ is set to 15 following typical settings. For each input BER, simulation is run until at least 10 frame errors have been collected to find the FER. This means, for example, more than $10^7$ decoding samples have been run for BER=0.025.}

\subsection{Ineffectiveness of existing logic locking}

{To find the effects of flipping signals by logic locking on the performance of LDPC decoding, experiments have been carried out in our work to flip one bit in every iteration of the Min-sum algorithm. Since the decoding results are hard decisions of the messages, the sign product, $s_m$, and the most significant bit (MSB) of $min1_m$ are among the signals that affect the decoding results most. Hence, one of the check nodes is chosen randomly, and its sign product or the MSB of the $min1$ value is flipped in every iteration in our simulations. The results are also shown in Fig. \ref{LDPC_comp}.} It can be observed that even if the sign product or the MSB of $min1$ of a check node is flipped in every iteration, the resulted differences in the FER and throughput are negligible. The reason is that the LDPC decoder is self-correcting. Even if a c2v message becomes wrong because of the flipping, the other c2v messages sent to the same variable node and the channel information are added up to compute the v2c messages and the {\it a posteriori} information as in Lines 17 and 20 of Algorithm \ref{Minsum}. Only the minimum and second minimum magnitudes of the v2c messages affect  later decoding iterations. As a result, the chance that the decoding becomes non-convergent or needs more iterations is very low. When the output of an existing SAT-attack-resistant logic-locking scheme is used to XOR with $s_m$ or MSB of $min1_m$, it cannot flip the signal in every LDPC decoding iteration for every input data when a wrong key is used. Hence, the resulted FER and throughput degradation would be even less significant compared to those shown in Fig. \ref{LDPC_comp}.

{It is also possible to XOR the output of an existing logic-locking block with multiple intermediate signals of the LDPC decoder. When there are too many flipped bits, they can not be corrected by LDPC decoding. However, in order to resist the SAT attack, each wrong key should only make the logic-locking block output `1' for a very small portion of all possible input patterns. For example, in the Anti-SAT design \cite{Anti-SAT}, a wrong key only makes the output `1' for one of the $2^n$ input patterns, where $n$ is the number of data inputs to the logic-locking block and needs to be reasonably large to resist the SAT attack. In this case, even if the output of the LDPC decoder is corrupted by using the wrong key, the degradation on the FER is still negligible.}

{In addition to the key bits used in the Anti-SAT or SARLock blocks, more key bits can be directly connected with intermediate signals by using AND/OR/XOR gates as shown in Fig. 1 in order to increase the corruptibility of the wrong keys. In this case, the FER of the LDPC decoder can be degraded significantly. However, those wrong key bits directly ANDed/ORed/XORed with signals can be derived by the AppSAT attack \cite{AppSAT}.}

{The FSM of an LDPC decoder is very simple. It consists of only three states. The decoder starts from the initial state. The second state is for iterative message computation. This state is repeated until the decoder stops and ends at the final state. Due to the simplicity of the FSM, it is very difficult to insert or bypass any states. Hence existing logic-locking schemes targeting at the FSM, such as HARPOON \cite{HARPOON} and parametric locking \cite{YasinPara}, would not work on LDPC decoders. Besides, they are subject to the SAT or resynthesis attack. }

\section{Secure Obfuscation Schemes for LDPC Decoders}
\begin{figure}
    \centering
    \includegraphics[width=3.0in]{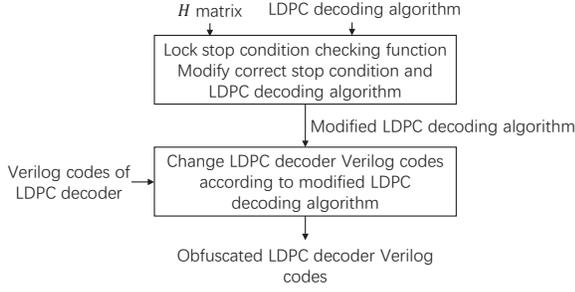}
    \caption{The design flow of the proposed LDPC decoders}
    \label{Design flow}
\end{figure}

In this section, two logic-locking methods incorporating the properties of the LDPC decoding algorithm are first proposed. By locking the stop condition checking function, which is the same in every LDPC decoding algorithm, our proposed designs can significantly degrade the throughput and/or error-correcting performance when the wrong key is applied. If the decoder can not achieve the FER or throughput required by the system, it becomes non-usable. To obscure the stop condition and effectively resist removal attacks, modifications to the decoding algorithms are also proposed in Subsection C. {The overall design flow is shown in Fig. \ref{Design flow}. For a given LDPC decoding algorithm, the stop condition checking function is locked, the correct stop condition and LDPC decoding algorithm is also modified. Then the Verilog codes of an LDPC decoder need to be changed according to the modification made on algorithmic level. In the end, an LDPC decoder obfuscated by the proposed obfuscation methods is the output.}



\subsection{Obfuscated LDPC decoder with throughput degradation at wrong keys}
\begin{figure}
    \centering
    \includegraphics[width=2.0in]{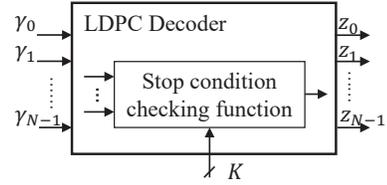}
    \caption{Obfuscated LDPC decoder by locking the stop condition checking function }
    \label{overall}
\end{figure}

The Min-sum decoding algorithm is iterative. Let $t=zH^T$. The decoding stops when $t=0$ and the most updated $z$ vector is the correct codeword. If $t$ does not become zero, the decoding runs until $I_{max}$ iterations, in which case decoding failure is declared. In the proposed design, as shown in Fig. \ref{overall}, the stop condition is controlled by a key vector and it is changed when a wrong key is used. Applying a wrong key leads to one of the two effects: 1) even though the correct codeword is already found and hence the $z$ vector does not change anymore, the decoding still runs until the last iteration; 2) the decoding process terminates before the correct codeword is found. The first case does not affect the error-correcting performance but reduces the throughput of the decoder. The second case increases the decoding failure rate. Hence, replacing the original stop condition checking function, ($t==0$), by a key controlled stop condition checking can effectively obfuscate LDPC decoders. Next, our first key-controlled stop condition logic-locking scheme is presented. 
\begin{figure}[t]
    \centering
    \includegraphics[width=1.5in]{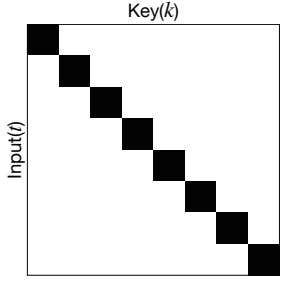}
    \caption{The K-map pattern of function $f_{1}(t,k)$}
    \label{Diag}
\end{figure}

The original stop condition is $t==0$. If $t$ has $h$ bits, the original stop condition checking function is 
\begin{equation}\label{ft}
    f(t) = \overline{t_0} \& \overline{t_1} \& \cdots \& \overline{t_{h-1}}.
\end{equation}
By adding a constant vector, $v$, decided by the designer and slightly modifying the LDPC decoding algorithm as will be detailed in subsection C of this section, the decoding converges when $t==vH^T$. Our first obfuscation method locks the stop condition checking function by $t == k$. It means that the right key is $vH^T$ and only when the vector $t$ equals the key input, $k$, the LDPC decoding stops. The Boolean expression of our logic-locking function is 
\begin{equation*}
    f_{1}(t,k) = \overline{(t_0 \oplus k_0)} \& \overline{(t_1 \oplus k_1)} \& \cdots \& \overline{(t_{h-1} \oplus k_{h-1})}
\end{equation*}
The K-map pattern of the $f_{1}(t,k)$ function is shown in Fig. \ref{Diag}. The black cells in Fig. \ref{Diag} mean that $f_1(t,k)$ is `1' for the corresponding key and input, in which case the LDPC decoding stops.

When the stop condition checking function is locked by $f_{1}(t,k)$, using a wrong key changes the stop condition. For a given wrong key, the decoding only stops when the syndrome $t$ equals the wrong key. Hence, most likely, the LDPC decoding will run to the last iteration. However, the original LDPC decoding will not run until the last iteration in most cases at the operating region. As shown for the example code in Fig. \ref{LDPC_comp}, even though $I_{max}$ is set to 15 to achieve good error-correcting capability, the decoding takes on average around 5 iterations to converge for the practical operating range of FER$<10^{-4}$. Hence the throughput of the obfuscated LDPC decoder is significantly decreased when a wrong key is used. {On the other hand, if a wrong key is applied, the obfuscated decoder output, $z$, is only different from the original decoding result when $t$ accidentally equals the wrong key before the decoding converges. As a result, the FER is almost the same.} In the case of obfuscated LDPC decoders, the corruptibility of a wrong key is defined as the number of input patterns leading to the wrong decoder output regardless of the number of decoding iterations. Hence, the wrong keys in our first proposed design all have low corruptibility. 

The size of the key input is equal to the number of rows in the $H$ matrix of the LDPC code in the above scheme. For practical LDPC codes, the $H$ matrix has hundreds or even thousands of rows. Using such a long key leads to high logic complexity and requires large memory. To address this issue, a shorter key input can be used to lock some of the bits in $t$. Assume that the key input is reduced to $h_k < h$ bits. Then the key-controlled function $f_1(t,k)$ can be modified to
\begin{equation}\label{update ob1}
    f_{2}(t,k) = \overline{(t_0 \oplus k_0)} \& \cdots \& \overline{(t_{h_k-1} \oplus k_{h_k-1})} \& \hat t_{h_k} \& \cdots \& \hat t_{h},
\end{equation}
where $\hat t_i=t_i$ or $\overline t_i$ when the corresponding bit in $vH^T$ is `1' or `0', respectively. The LDPC decoding stops when the first $h_k$ bits of $t$ equal the key input bits and the other bits are the same as those in $vH^T$. 

{In summary, the first proposed logic-locking scheme for LDPC decoder can be designed according to the following steps:
\begin{enumerate}
    \item Select the number of key bits $h_k$ and choose a random secret vector $v$.
    \item Design the stop condition checking function as $f_2(t,k)$ according to \eqref{update ob1}.
    \item The correct key $k^*$ equals to the first $h_k$ bits of $vH^T$.
\end{enumerate}
Such a logic-locking block can be inserted in the behavioral description of the LDPC decoder.}

{All the wrong keys in this obfuscated LDPC decoder have low corruptibility as one. Each of the wrong keys degrades the decoder throughput and has negligible effect on the error-correcting performance in the same way. They are not affected by the key length $h_k$. However, $h_k$ should be large enough so that the right key can not be easily guessed.}

\subsection{Obfuscated LDPC decoder with error-correcting performance and throughput degradation at wrong keys}
In our first obfuscation method, all wrong keys are of low corruptibility and the resulted error-correcting performance degradation is negligible. In this subsection, our second design with high-corruptibility wrong keys is proposed. The high-corruptibility wrong keys make the decoder prematurely stop at multiple values of the $t$ vector before the codeword is found. As a result, the error-correcting performance of the obfuscated LDPC decoder is significantly degraded when a high-corruptibility wrong key is used. Besides, similar to the first design, a wrong key leads to substantial reduction on the decoding throughput since the decoding will run until the last iteration in most cases.

The decoder input hard decision is $z = c\oplus e$, where $c$ and $e$ are the codeword and error vector, respectively. For practical channel BERs, the number of errors happened is small. Hence, $t=zH^T=(c \oplus e)H^T = eH^T$ has a limited number of patterns and the patterns are not randomly distributed. Take the (1270, 635) LDPC code as an example. If the FER needs to be lower than $10^{-6}$, then the BER should be less than 0.025. For this code with row weight $d_c=10$, it can be computed that the probability of each bit of $t$ being `1' is $\sum_{i=1, i\text{ is odd}} ^{i\leq d_c} \binom{d_c}{i}\times 0.025^i \times 0.975^{d_c-i} \approx 0.201$. Hence, $t$ has lower Hamming weight in most of the cases. If $t$ is directly used as the input of the logic-locking block and a random wrong key is picked, then most likely the wrong key will not make the decoding stop prematurely and the error-correcting performance of the decoder will not change much. 

To substantially degrade the LDPC decoder error-correcting performance, the vector $t$ needs to be mapped to a randomly-distributed vector, $r$, to be used as the input of the logic-locking block. In other words, the mapping should be designed such that $Pr\{r_i=1\} \approx 0.5$. Divide the $t$ vector into groups of $g$ bits. In our design, 
\begin{equation}\label{map}
r_i = t_{i\times g} \oplus t_{i\times g + 1} \oplus \cdots \oplus t_{i\times g+g-1}.
\end{equation}
Besides, $g$ is chosen to make $Pr\{r_i=1\} \approx 0.5$. Take the (1270, 635) LDPC code in Fig. \ref{LDPC_comp} as an example. When BER=$0.025$, $Pr\{r_i = 1\}$ can be calculated as $\sum_{j=1, j\text{ is odd}}^{j\leq g}{g \choose j}\times 0.201^{j}\times (1-0.201)^{g-j}$. 
$Pr\{r_i = 1\}$ is around $0.5$ if $g=15$ is used. For a different BER and/or a LDPC code with different row weight, $g$ can be decided in a similar way. Changing the stop condition to a nonzero vector as will be detailed in the next subsection does not affect this mapping function or the value of $g$.

\begin{figure}[t]
    \centering
    \includegraphics[width=3.3in]{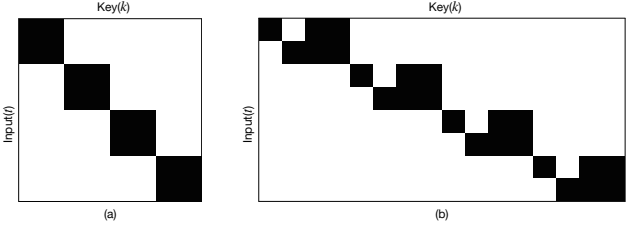}
    \caption{(a) The K-map of the $f_3$ function with only high-corruptibility wrong keys; (b) the K-map of the $f_4$ function}
    \label{Diag_update}
\end{figure}

Assume that the $l_r$ bits of $r$ are generated from $t$ through the mapping. 
If the logic-locking block is to make the decoding stop when the key equals the input as in our first proposed scheme, then its function making use of $r$ is 
\[f_{3}(r,k)= \overline{(r_0 \oplus k_0)} \& \cdots \& \overline{(r_{l_r-1} \oplus k_{l_r-1})}.\]
Each key makes the decoding stop at one pattern of $r$.  If the number of rows in the parity check matrix of the LDPC code is $h$, each pattern of $r$ corresponds to $2^{h-l_r}$ patterns of $t$. Therefore, as shown in the K-map of the $f_{3}$ function in Fig. \ref{Diag_update}(a), each wrong key has high corruptibility and makes the decoding stop at $2^{h-l_r}$ different patterns of $t$. The probability that the LDPC decoding process stops in each iteration at such wrong keys is upper bounded by $1/2^{l_r}$ and it most likely decreases over the iterations. Since the FERs required for practical applications are low, such as less than $10^{-5}$ or $10^{-6}$, the wrong keys lead to significant increase in the FERs. Smaller $l_r$ degrades the FER more. On the other hand, the wrong keys with higher corruptibility are more likely to be excluded within a small number of iterations in the AppSAT attack \cite{AppSAT}. A proper value needs to be picked for $l_r$ to balance the FER degradation and AppSAT attack resistance. Besides, the decoding always runs until the last iteration when $r$ does not equal the wrong key. Hence, the high-corruptibility wrong keys also reduce the throughput of the LDPC decoder.

If all wrong keys have high corruptibility as shown in Fig. \ref{Diag_update}(a), the SAT resistance cannot be guaranteed. Wrong keys with corruptibility as one need to be introduced to resist the SAT attack \cite{G-Anti-SAT}. Each wrong key for function $f_2$ in \eqref{update ob1} makes the LDPC decoding stop at one pattern of $t$. Such a function can be utilized to introduce wrong keys of corruptibility one. The logic-locking function of our second obfuscation scheme is
\begin{equation}\label{general express}
    f_4(t,k) =  f_3(r,kb) \& (f_2(t,ka) + f_h(r,ka)),
\end{equation}
where $k = [kb||ka]$ and ``$||$'' denotes concatenation. Similarly, $kb$ has $l_r$ bits. Since the $l_r$ bits of $r$ are generated from $gl_r$ bits of $t$, the key length of $ka$ is set to $gl_r$. $ka$ can be also made longer at the cost of higher logic complexity and larger memory for storing the key. In \eqref{general express}, $f_h(r,ka) = (r_0 \oplus ka_{0} \oplus \cdots \oplus ka_{g-1}) +\cdots+  (r_{lr-1} \oplus ka_{g(l_r-1)} \oplus \cdots \oplus  ka_{gl_r-1})$ is ANDed with $f_3$ to make the high-corruptibility wrong keys of $f_4$ have the same corruptibility as the wrong keys of the $f_3$ function. The K-map of $f_4$ is shown in Fig. \ref{Diag_update}(b). By ANDing $f_3(r,kb)$ with $f_2(t,ka)$, $f_4=$ `$1$' in the cells of the K-map whose labels satisfy that $kb=r, ka=\ [t_0 \cdots t_{gl_r-1}]$, and the other bits of $t$ equal the corresponding bits of $vH^T$. These cells are the single black cells in the respective columns in Fig. \ref{Diag_update}(b) and their column labels are the low-corruptibility keys. The $f_3(r,kb) \& f_h(r,ka)$ in \eqref {general express} makes $f_4=$`$1$' in the cells whose labels satisfy that $kb = r,\ [(ka_{0} \oplus \cdots \oplus ka_{g-1}), \cdots ,(ka_{g(l_r-1)} \oplus \cdots \oplus  ka_{gl_r-1})] \neq r$, and the column labels for these cells are the high-corruptibility keys. 

{In summary, the second proposed logic-locking scheme for LDPC decoder can be designed according to the following steps:
\begin{enumerate}
    \item Select the values of $l_r$ and $g$; Choose a random secret vector $v$. 
    \item Design the stop condition checking function as $f_4$ according to \eqref{general express}.
    \item Map the first $g\times l_r$ bits of vector $vH^T$ to an $l_r$-bit vector $r^*$ using a formula similar to that in \eqref{map}. The correct key is $k^*=[kb^*||ka^*]$, where $kb^*=r^*$ and $ka^*$ equals the first $g\times l_r$ bits of vector $vH^T$.
\end{enumerate}}

Using the $f_4$ function, the keys are divided into three categories: 1) a single correct key that makes the LDPC decoding stop at $t==vH^T$; 2) the low-corruptibility wrong keys that reduce the throughput but affect the error-correcting performance of the LDPC decoder negligibly; 3) the high-corruptibility wrong keys that degrade both the error-correcting performance and throughput. The ratio between the high-corruptibility and low-corruptibility wrong keys is affected by $l_r$. When $l_r$ is larger, the ratio is larger. However, the high-corruptibility wrong keys have lower corruptibility value and accordingly they result in less significant degradation on the error-correcting performance. Hence, $l_r$ allows a tradeoff between the corruptibility of the wrong keys and portion of high-corruptibility wrong keys. Nevertheless, even if a low-corruptibility wrong key is used, there is significant degradation on the throughput and the decoder becomes not usable. The degradation on the decoder throughput is not much dependent on $l_r$.


\subsection{LDPC decoding algorithm with modified stop condition}
The main idea of the two proposed obfuscation methods is to lock the stop condition checking function by a key and the correct key equals the syndrome when the decoding converges. The stop condition for the original LDPC decoding is $t==0$. A potential threat is that the zero stop condition can be easily guessed by an attacker who has knowledge about LDPC decoders. If the stop condition checking logic-locking block can be identified from the netlist, then it can be eliminated and replaced by a zero detector. Next, low-overhead modifications to the decoding algorithm are proposed to address this issue. 
\begin{algorithm}[t]
\caption{The Modified Min-sum Decoding Algorithm}
\label{modified Minsum}
\begin{algorithmic}[1]
\STATE { {\textbf{input:} $\gamma_n$}}
\STATE {\textbf{initialization:}$|\gamma'_n|\!=\! |\gamma_n|;\mathbf{z_n'\!=\! sign(\gamma'_n) \!=\! sign(\gamma_n) \oplus v_n}$};\\  \indent\hspace{6em} $u_{m,n}' = \gamma_n'$
\STATE{if $\mathbf{z'H^T = p}$}
\STATE{\quad {return codeword $\mathbf{z=z'\oplus v}$ and stop}}
\STATE{for $k = 1$ to $I_{max}$}
\STATE {\quad \textbf{check node processing}}
\STATE {\quad {for each $m$}}
\STATE {\quad \quad $min1_m' = min_{j\in S_v(m)}|u_{m,j}'|$}
\STATE {\quad \quad $idx_m' = argmin_{j\in S_v(m)}|u_{m,j}'|$}
\STATE {\quad \quad $min2_m' = min_{j\in S_v(m),\ j\neq idx_m'}|u_{m,j}'|$}
\STATE {\quad \quad $s_m' = \prod_{j\in S_v(m)}sign(u_{m,j}')$}
\STATE{\quad \quad for each $n \in S_v(m)$}
\STATE {\quad \quad \hspace{2em} $|v'_{m,n}| =  \left\{
             \begin{array}{lr}
             \alpha min1_m'\ if\  n \neq idx_m' &  \\
             \alpha min2_m'\ if\  n = idx_m' & \\  
             \end{array}
\right.$}
\STATE {\quad \quad \hspace{2em} $sign(v_{m,n}') = \mathbf{(-1)^{p_m}}s_m' sign(u_{m,n}')$}
\STATE{\quad \textbf{variable node processing}}
\STATE {\quad {for each $n$ and $m$}}
\STATE{\quad \quad $u_{m,n}' = \gamma_n' + \sum_{i\in S_c(n),i\neq m}v_{i,n}'$}
\STATE{\quad \textbf{a posteriori info. comp. \& tentative decision}}
\STATE {\quad {for each $n$}}
\STATE{\quad \quad $\Tilde{\gamma_n'} = \gamma_n' + \sum_{i\in S_c(n)}v_{i,n}'$}
\STATE{\quad \quad $z_n' = sign(\Tilde{\gamma_n'})$}
\STATE {\quad  {if $\mathbf{z'H^T = p}$ }}
 \STATE {\quad\quad {output codeword $\mathbf{z=z' \oplus v}$ and stop}}
\STATE{end for}
\STATE{ {output $\mathbf{z =z'\oplus v}$ and declare decoding failure}}
\end{algorithmic}
\end{algorithm}

In our design, a constant vector, $v$, chosen by the designer is added to the input vector $z=c\oplus e$. Hence the decoding is carried out on $z' = c\oplus e\oplus v$. The decoding algorithm can be modified so that it generates $c\oplus v$ when the error vector, $e$, is decodable. Then $c$ is recovered by adding $v$ back. Such modifications lead to exactly the same decoding results and do not cause any error-correcting performance loss. However, in this case, the correct stop condition is changed to whether $t$ equals $p=vH^T$, which is not an all-`0' vector. To incorporate the $v$ vector, the Min-sum decoding can be modified as in Algorithm 2. At initialization, $v_n$ is added to the sign of $\gamma_n$ to generate the initial hard-decision vector $z'$ and $|\gamma_n'|=|\gamma_n|$. Accordingly, the initial $u'_{m,n}$ equals $\gamma'_n$. In the first iteration, $min1_m',\ idx_m'$, and $min2_m'$ are computed in the same way and they equal $min1_m,\ idx_m$, and $min2_m$, respectively, in Algorithm 1, since the magnitudes of the channel information are not changed. Also $(-1)^{p_m} s'_m$ used in Line 14 of Algorithm 2 equals the $s_m$ in Algorithm 1. The signs of all the messages added up in the variable node processing and {\it a posterior} information computation in Algorithm 2 are different from those in Algorithm 1 by $v_n$. As a result, the $z'$ vector in Algorithm 2 and the $z$ vector in Algorithm 1 is always different by $v$ in every iteration. Hence, by adding $v$ back after the decoding, the same codeword is recovered.


The vector $v$ is secret. However, it does not need to be stored
in memory. The additions of $v$ to the LDPC decoder input and output and the multiplication of $(-1)^{p_m}$ to the sign can be implemented by adding NOT gates to the inputs and outputs of the corresponding variable node units (VNUs) that compute v2c messages and hard decisions using c2v messages and channel inputs. The added NOT gates are merged with the other logic gates by the synthesize tool and can not be told from the netlist. Besides, the VNUs are interconnected with other decoder components through barrel shifters, which route the v2c and c2v messages according to the nonzero entries of the $H$ matrix. Since all the decoder components are also mixed by the synthesis tool, individual VNU cannot be isolated from the netlist. As a result, it is very difficult for the attacker to tell which VNUs have NOT gates added and accordingly recover the $v$ vector. {Also adding NOT gates does not necessarily increase the area or critical path since inverting gates, such as NAND and NOR, are smaller and faster than non-inverting gates, such as AND and OR \cite{DigitalBook}}. 

\begin{figure}
    \centering
    \includegraphics[width=3.5in]{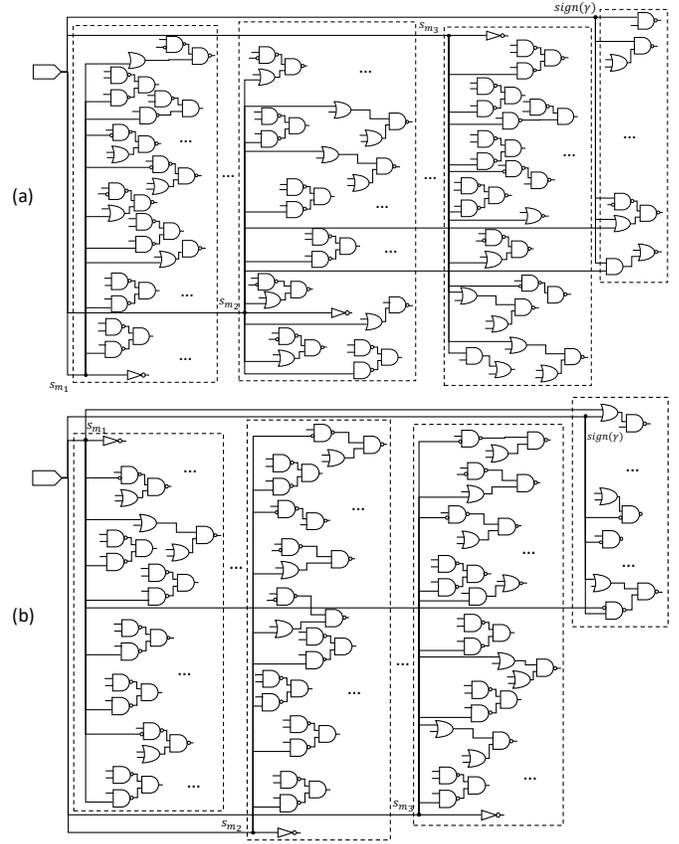}
    \caption{Schematic illustration of a VNU with three c2v input messages for LDPC decoding synthesized using TSMC $65nm$ process under $4ns$ timing constraint with (a) the signs of the messages from the first and third check nodes flipped; (b) the signs of channel input and messages from the first and second check nodes flipped.}
    \label{synth_fig}
\end{figure}

To show that the NOT gates inserted to incorporate the $v$ vector addition and $(-1)^{p_m}$ multiplication are merged with the other gates by the synthesis tool, two experiments are carried out on a toy VNU with three input c2v messages, whose sign bits are denoted by $s_{m_1}$, $s_{m_2}$, and $s_{m_3}$,  and one channel information input, denoted by $\gamma$, using TSMC $65nm$ process with $4ns$ timing constraint. The synthesis results are translated to logic diagrams and the beginning parts connected to the flipped inputs are shown in Fig. \ref{synth_fig}. In the first experiment whose result is shown in part (a) of this figure, $s_{m_1}$ and $s_{m_3}$ are flipped due to the multiplication of $(-1)^{p_m}$ in Line 14 in Algorithm \ref{modified Minsum} while $s_{m_2}$ is not. In the second experiment whose result is illustrated in Fig. \ref{synth_fig}(b), $s_{m_1}$ and $s_{m_2}$ are flipped while $s_{m_3}$ is not. Besides, the sign bit of $\gamma$ is flipped due to the addition of the $v$ vector according to Line 2 in Algorithm \ref{modified Minsum}. From this figure, the inputs with NOT gates inserted can not be differentiated from those without inserted NOT gates. Therefore, neither the $v$ or $p=vH^T$ vector can be derived from the netlist. Considering that the VNU is also mixed with the other computation units in the LDPC decoder by the synthesis tool, it is even harder for the attacker to get $v$ or the correct stop condition. 

\begin{table}[h]
    \centering
    \caption{{Areas and power consumption of VNUs synthesized using TSMC $65nm$ process with $4ns$ timing constraint}}
    \begin{tabular}{c||c|c}
         \hline
          & Area $(\mu m)^2$  & Power $(nW)$\\
         \hline
         First modified VNU &  381.96 & 67295.42 \\
         \hline
         Second modified VNU & 380.88 & 70329.78 \\
         \hline
         Original VNU & 380.52 & 67191.04 \\
         \hline
    \end{tabular} 
    \label{toy comparision}
\end{table}

{The area and power consumption of the two modified VNUs shown in Fig. \ref{synth_fig} and the VNU without any modification reported by the synthesis tool are listed in Table \ref{toy comparision}. It can be observed that the first and second modified VNUs have 0.38\% and 0.09\%, respectively, area overheads compared to the original VNU. Their power consumption is only 0.16\% and 4.7\% higher compared to that of the original VNU. Considering that an LDPC decoder also has large memory and other computation units, the area and power consumption overheads brought to an LDPC decoder by the NOT gate insertions is even smaller. }


QC-LDPC codes \cite{QC-LDPC} are usually used in practical systems since they enable efficient parallel processing. The $H$ matrix of a QC-LDPC code consists of sub-matrices, each of which is either a zero or a cyclically shifted identity of dimension $q\times q$. Many parallel LDPC decoder designs process one sub-matrix or one block column of sub-matrices in each clock cycle. In this case, $v$ can be a vector with repeated patterns of length $q$. Its additions can be also implemented by adding NOT gates.

\section{Experimental Results}
This section analyzes the security, effectiveness, and hardware complexity overheads of the proposed LDPC decoder obfuscation schemes. 

\subsection{SAT and AppSAT attack resistance}
{The SAT and AppSAT attacks can only be applied to combinational circuits. LDPC decoding algorithms are iterative. Registers are needed to store intermediate results, and they cannot be converted to CNF formulas. In order to apply the SAT/AppSAT attack, copies of the combinational parts of the LDPC decoder can be connected to eliminate the registers as in the unrolling technique \cite{unroll}. To unroll an LDPC decoder with $I_{max}$ decoding iterations for a code whose $H$ matrix has $d_c$ `1's in each row,  $I_{max}\times d_c$ copies of the combinational units are needed.} The decoder for practical LDPC code is already large. Hence it is infeasible to unroll the decoder for $I_{max}\times d_c$ times and carry out the SAT or AppSAT attack experiments for practical LDPC codes. In this subsection, mathematical analyses are first carried out on the resistance of the proposed designs to the SAT attack. Then SAT attack experiments are carried out for a toy LDPC decoder. After that, analyses on AppSAT attack resiliency are provided.

The main idea of the SAT attack is to exclude all wrong keys that make the obfuscated circuit output different from the original circuit output. For LDPC decoders, whether an output equals the original output is determined based on the returned vector, $z'$, without taking into account the number of decoding iterations. Unlike in the case that the decoder is operating at practical BER range and has a limited number of errors in the input vector, the SAT attack can utilize any input pattern. There are two types of input patterns for the LDPC decoder. The first type is the decodable inputs that make the LDPC decoding process converge at iteration $I\leq I_{max}$. The second type is the undecodable inputs that make the LDPC decoding run to the last iteration without convergence. The low-corruptibility wrong keys in both of the proposed obfuscated LDPC decoders make the decoding stop at one specific syndrome. Hence, when a decodable input pattern is selected as a DIP, the low-corruptibility wrong keys making the LDPC decoding process stop before iteration $I$ will be excluded. Therefore, such a DIP can exclude at most $I_{max}$ low-corruptibility wrong keys. When an undecodable input pattern is selected as a DIP, the decoding process will run until the last iteration. In this case, the low-corruptibility wrong keys that make the LDPC decoding process stop before the last iteration will be excluded. In the worst case, the syndrome vector $t$ is different in each decoding iteration. Therefore, at most $I_{max}$ low-corruptibility wrong keys will be excluded. As a result, for either case, a DIP can exclude at most $I_{max}$ low-corruptibility wrong keys. For the first and second proposed obfuscation designs, the numbers of low-corruptibility wrong keys are $2^{h_k}$ and $2^{gl_r}$, respectively. 
{Hence the minimum number of queries needed to precisely recover the functionality in the two proposed obfuscated LDPC decoders are $2^{h_k}/(I_{max})$ and $2^{gl_r}/(I_{max})$. Both of them are exponential to the number of key bits, and hence the SAT attack is effectively resisted.}


\begin{table}[t]
    \centering
    \caption{SAT attack experimental results on (56, 28) LDPC decoders}
    \begin{tabular}{c||c|c|c|c}
         \hline
          & & $h_k$ = 7 & $h_k$ = 14 & $h_k$ = 21 \\
         \hline
         \multirow{2}{2.2cm}{The first obfuscated decoder} & \# of iterations & 128 & 16384 & - \\
         \cline{2-5}
         & time (second) & 1.21 & 12137.5 & - \\
         \hline
          & & $gl_r$ = 7 & $gl_r$ = 14 & $gl_r$ = 21 \\
         \hline
         \multirow{2}{2.2cm}{The second obfuscated decoder} & \# of iterations & 128 & - & - \\
         \cline{2-5}
         & time (second) & 3.32 & - & - \\
         \hline
    \end{tabular}
    
    \label{comparison SAT}
\end{table}

To further evaluate the SAT-attack resiliency of our proposed designs, the obfuscated LDPC decoders for a (56, 28) toy code is unrolled for one iteration, which is equivalent to setting $I_{max}=1$. Open-source SAT attack tool \cite{SAT tool} is applied on these toy decoders using an Intel Core i7 with 4GB RAM platform and the CPU time is limited to 10 hours. Table \ref{comparison SAT} shows the results. For the first obfuscation scheme, the SAT attack can be finished within 10 hours when its key length, $h_k$, is 7 or 14. For both of these key lengths, the number of iterations needed by the SAT attack is $2^{h_k}$, which matches our analyses for the case of $I_{max}=1$. For the second obfuscation design, the SAT attack can not be finished within 10 hours when $gl_r$ is 14 or larger. This design leads to longer SAT attack time mainly because it has more complicated logic compared to the first obfuscation scheme. The SAT attack finished after $128=2^7$ iterations in the case of $gl_r=7$. This result also matches our analyses for the second obfuscation design.

All the wrong keys for the first proposed obfuscated decoder have low corruptibility.
{When the $H$ matrix of the LDPC code is $(N-S)\times N$, the vector $t$ has $N-S$ bits. If the AppSAT attack is applied, any returned low-corruptibility wrong key would allow the attacker to recover the approximate functionality with $\epsilon =1/(2^{N-S})$ \cite{approximation}, which means that one out of the $2^{N-S}$ input patterns leads to different outputs.}
However, a low-corruptibility key most likely makes the decoder run until the last iteration and leads to significant decrease in the throughput. The ratio between the high-corruptibility and low-corruptibility wrong keys in the second proposed design is around $2^{lr}$. Although high-corruptibility wrong keys are more likely to be excluded by the AppSAT attack, some high-corruptibility wrong keys are left in the pool. If a high-corruptibility wrong key is returned by the AppSAT attack and is used for the decoder, it will lead to degradation on not only the throughput but also the FER of the decoder. If a low-corruptibility wrong key is returned, even though it can help to recover the approximate functionality of LDPC decoder, the throughput is degraded. For either case, the returned key is not usable. 


\begin{figure}
    \centering
    \includegraphics[width=3.5in]{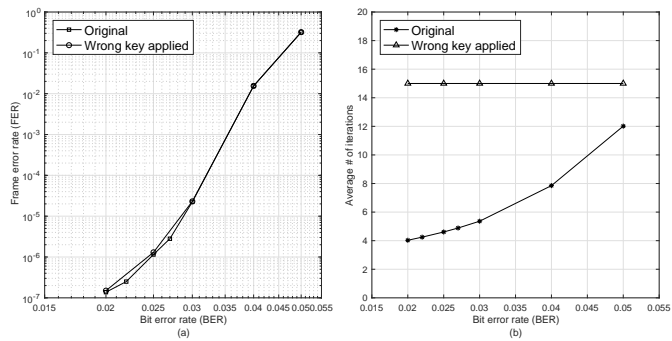}
    \caption{Min-sum decoding results for (1270, 635) QC-LDPC code with $I_{max}=15$ using the first proposed obfuscation scheme that only has low-corruptibility wrong keys: (a) frame error rate; (b) average number of iterations}
    \label{option1 sim}
\end{figure}
\begin{figure}
    \centering
    \includegraphics[width=3.5in]{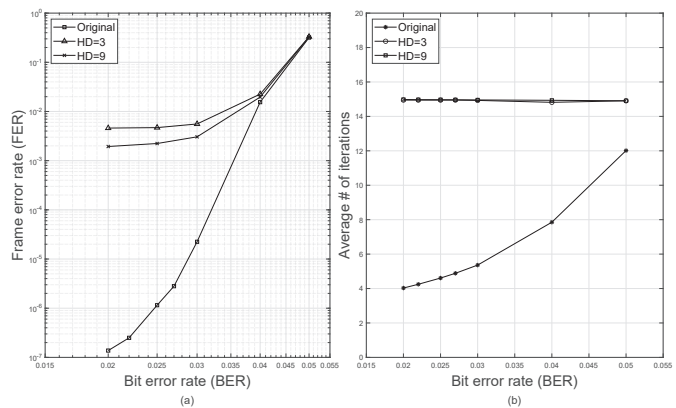}
    \caption{Min-sum decoding results for (1270, 635) QC-LDPC code with $I_{max}=15$ using the second proposed obfuscation scheme that has both high and low-corruptibility wrong keys: (a) frame error rate; (b) average number of iterations}
    \label{option2 sim}
\end{figure}

\subsection{Effectiveness of the proposed obfuscation schemes}

To show the performance degradation that can be brought by the proposed obfuscation schemes to the LDPC decoder when low and high-corruptibility wrong keys are used, simulations are carried out using C++ for the example (1270, 635) QC-LDPC code, whose $H$ matrix consists of $5\times 10$ sub-matrices of dimension $127\times 127$. $I_{max}$ is set to 15 in our simulations.

Simulation results for the first obfuscated LDPC decoder whose wrong keys are all low-corruptibility are shown in Fig. \ref{option1 sim}. {In our simulations, the key length, $h_k$, is set to 127. A 127-bit random vector that is not equal to the first 127 bits of $vH^T$ is used as the low-corruptibility wrong key.} As illustrated in part (b) of this figure, in the original LDPC decoder, the average number of decoding iterations is much smaller than $I_{max}$. However, if a wrong key is used, the decoder most likely runs until the last iteration and the average number of iterations becomes almost $I_{max}$. For the practical operating range of BER$<$0.03, this translates to more than three times reduction on the throughput. On the other hand, each wrong key makes the decoder stop at one pattern of $t$, and the chance that the decoding stops prematurely at a wrong vector is small. As a result, as shown in Fig. \ref{option1 sim}(a), the first proposed obfuscation method only leads to negligible difference in the FER. 

In the second proposed design, if a low-corruptibility wrong key is used, the decoder will have significant degradation on the throughput but negligible difference in the FER similar to those in Fig. \ref{option1 sim}. The simulation results when high-corruptibility wrong keys are used are shown in Fig. \ref{option2 sim}. {In these simulations, two high-corruptibility wrong keys with $lr=10$ and $g=15$ are used. The $lr$-bit $kb$ parts of the two keys are random vectors with Hamming distance (HD) equal to 3 and 9 compared to $r^*$. The  $lr\times g$=150-bit $ka$ part is randomly generated.} The corruptibilities of these two wrong keys are both $2^{h-l_r}=2^{625}$. The one with smaller HD causes more degradation on the FER as shown in Fig. \ref{option2 sim}(a). The reason is that the number of errors decreases and the HD between $t$ and $vH^T$ becomes lower over the LDPC decoding iterations. Hence, when a wrong key with smaller HD is used, the chance that it equals the $r$ vector and makes the decoding stop prematurely is higher compared to the case that a key with higher HD is utilized. Nevertheless, even for the almost largest possible HD=9 that can be used when $l_r=10$, the wrong key leads to more than two orders of magnitude degradation on the FER at practical operating range of BER$<$0.03. The probability that a high-corruptibility wrong key makes the decoding stop prematurely is also relatively small and hence the decoding also runs until the last iteration in most cases. Therefore, as shown in Fig. \ref{option2 sim}(b), the average number of iterations is also around $I_{max}$ when high-corruptibility wrong keys are used. The average iteration number for the HD=3 wrong key is slightly smaller compared to that of the case with HD=9, although the difference is not visible from the figure.

\subsection{Removal attack resistance}
 
Both of the proposed schemes obfuscate LDPC decoders utilizing algorithmic properties instead of just adding a logic-locking block to the circuit. It is possible that the block or part of the block implementing the function $f_2$ or $f_4$ can be identified, for example, by sorting the probability skew values of the signals \cite{removal}, since it involves large AND trees. However, these functions check the stop condition and they are integral parts of the decoder. If they are removed, the LDPC decoder will run until the last iteration and the throughput is substantially reduced. Additionally, by modifying the decoding algorithm with negligible complexity overhead, the stop condition is changed from an all-`0' vector to a random-like vector. This makes it impossible to replace the proposed logic-locking block by the correct function even if the attacker knows that a normal LDPC decoder stops at all-`0' syndrome vector. Besides, the NOT gates inserted for XORing the $v$ vector for the algorithmic modifications are combined with the other logic gates by the synthesis tool. Hence, the $v$ vector and accordingly the modified stop condition can not be recovered from the netlist either as analyzed in Section III. C. 


\subsection{Hardware overhead analyses}
This subsection analyzes the complexity overheads of the proposed obfuscation schemes brought to LDPC decoders by using the (1270, 635) code as an example. {The decoder with the proposed stop condition checking function for logic locking are described using Verilog codes in register transfer level. All of the synthesis results are provided using the Cadence Genus synthesis solution with TSMC $65nm$ process under $4ns$ timing constraint.}

The hardware implementation architectures of LDPC decoders depend on not only the decoding algorithm but also the computation scheduling scheme. Sliced message passing is a popular scheduling scheme due to its high throughput and low memory requirement \cite{MyBook}. In such a decoder, one block column of the $H$ matrix is processed in each clock cycle. Hence, $5\times 127 = 635$ CNUs are used to compute the $min1$, $min2$, {\it etc.} and then derive the c2v messages for each row of the $H$ matrix in parallel. Also 127 VNUs are used to compute the v2c messages and the {\it a posteriori information}. Besides, $2\times 5$ 127-input Barrel shifters are needed to route the messages between the CNUs and VNUs according to the nonzero entries of the shifted-identity sub-matrices in $H$. Details about the CNU and VNU can be found in \cite{MyBook}. {The components of the sliced-message-passing LDPC decoder are described using Verilog codes in register transfer level and synthesized by the Cadence Genus tool using TSMC 65$nm$ process. The areas reported by the tool under 4$ns$ timing constraint are listed in Table \ref{comparison area}.}
\begin{table}[h]
    \centering
    \caption{Areas of sliced-message-passing (1270, 635) LDPC decoders synthesized using TSMC $65nm$ process with 4$ns$ timing constraint}
    \begin{tabular}{c||c}
         \hline
          & Area $(\mu m)^2$  \\
         \hline
         CNUs &  93040.2 \\
         VNUs &  81280.0 \\
         routing networks &  81489.6 \\
         stop condition checking block & 536.4 \\
         \hline
         Original decoder & 256346.2 \\
         \hline
         \hline
         Overhead of first obfuscated decoder &  256732.8 (0.15\% overhead) \\\hline
         Overhead of second obfuscated decoder & 257766.0 (0.55\% overhead) \\
         \hline
         {Overhead of Anti-SAT block} & {257292.92 (0.36\% overhead)} \\\hline
         \hline
    \end{tabular} 
    
    \label{comparison area}
\end{table}

The first obfuscation scheme uses the $f_2(t,k)$ function in \eqref{update ob1} to implement the stop condition checking. Compared to the original stop condition checking function $f(t)$ in \eqref{ft}, the overheads include $h_k$ XOR gates, where $h_k$ is the number of key bits. Even if $h_k$ is set to 127 to achieve high security level, the proposed obfuscation scheme only brings $0.15\%$ overhead to the area. Besides the computation units listed in Table \ref{comparison area}, large memories are needed to store the channel input of the decoder and the hard-decision vector. Hence, the overhead brought by the proposed scheme to the overall decoder area is even smaller. 

The $f_2(t,k)$ function has one more gate in the data path compared to $f(t)$. However, the VNUs have several multi-bit adders and two sign-magnitude to 2's complement converters in the data path. It is much longer than the data path of the stop condition checking and decides the critical path of the LDPC decoder. Therefore, the proposed scheme does not cause any degradation on the achievable clock frequency of the decoder.

The second obfuscation scheme utilizes the $f_4(t,k)$ function in \eqref{general express} for stop condition checking. When the values of $g$ and $l_r$ are set to 15 and 10, respectively, the sizes of $ka$ and $kb$ are 10 and 150 bits, respectively. This design leads to $0.55\%$ area overhead compared to the original decoder. Although this overhead is larger than that of the first obfuscation design, it is still very small.

{For comparison, the Anti-SAT block is also synthesized and the results are included in Table \ref{comparison area}. In order to achieve a similar number of iterations needed by the SAT attack as the first proposed obfuscation scheme with $h_k=127$, the key length of the Anti-SAT design is set to 254-bit. In this case, the area overhead brought by the Anti-SAT design is 0.36\%, which is larger than that of the first obfuscation scheme but smaller than that of the second one. However, the Anti-SAT design is only to flip signals in conventional designs of LDPC decoders without the proposed stop condition checking function modification. Even if a wrong key is used, it would not lead to any noticeable performance loss as analyzed in Section II.D.}

\section{Discussions}
{This paper takes the Min-sum decoding algorithm as an example to illustrate the proposed logic-locking schemes, since this algorithm is used in many applications. There are other LDPC decoding algorithms. They are different from the Min-sum algorithm in the way that the magnitudes of the messages are computed in the check and variable node processing. However, they all have the same stop condition and sign computation as the Min-sum algorithm. The proposed obfuscation schemes modify the stop condition and sign computation. Hence, they can be applied in the same way to other LDPC decoders implementing different decoding algorithms. }

Using an existing logic-locking scheme, such as the Anti-SAT, G-Anti-SAT, DTL, or SFLL design, to lock one bit of $t$ does not effectively corrupt the LDPC decoder output. Flipping one bit of $t$ makes the LDPC decoding stop at a wrong pattern of $t$. However, the chance that $t$ equals a pattern that has a single bit different from $vH^T$ is very low. Hence, such logic-locking method does not lead to any noticeable increase in the FER. Besides, the outputs of existing lock-locking blocks are not always `1'. Even if a scheme with relatively high corruptibility, such as the DTL design \cite{AppSAT}, is used, the selected bit of $t$ will not be flipped with high probability. Hence, even if the decoder stops at a later iteration, only one or two more iterations are carried out instead of until the last iteration. Accordingly, existing logic-locking schemes do not cause significant degradation on throughput or FER as the proposed designs.

Another possible method to obfuscate the LDPC decoder is to XOR the output of an existing logic-locking block directly with one bit in the decoder output vector. However, by comparing the output of the locked decoder with that of a functioning decoder, the location of the flipped bit can be easily identified and the logic-locking block is subject to removal attacks. {The SFLL-FLEX scheme \cite{SFLL} has not been broken by any removal attacks. Nevertheless, there is a trade-off between the degradation it can cause on the LDPC decoder FER and the resistance to the SAT attack. For the SFLL-FLEX block that achieves high resistance to the SAT attack, XORing its output with LDPC decoder output will only lead to negligible FER degradation when a wrong key is used. Besides, XORing the output of any logic-locking block with the LDPC decoder output does not change the number of decoding iterations and hence does not bring any degradation to the throughput.} 

\section{Conclusions}
In this paper, two obfuscation schemes for LDPC decoders have been proposed by locking the stop condition checking function and modifying the decoding algorithm. Both designs have low-corruptibility wrong keys that make the SAT attack complexity exponential. Also they cannot be all excluded by the AppSAT attacks. These wrong keys lead to significant degradation on the decoder throughput. In addition to the low-corruptibility wrong keys, there are a large portion of high-corruptibility wrong keys in the second proposed obfuscation method. They cause significant degradations on not only the throughput but also the error-correcting performance. Besides, the proposed modifications on the decoding algorithm make the correct decoding stop condition a secret and thwart possible attacks. Future work will study algorithmic obfuscations on other fault-tolerating or self-correcting functions.

\begin{IEEEbiography}{Jingbo Zhou}
received the B.S. degree in telecommunication engineering from Beijing University of Post and Telecommunication, Beijing, China. He is currently pursuing the Ph.D. degree in the Electrical and Computer Engineering Department, The Ohio State University, OH, USA.

His current research interest is hardware security.
\end{IEEEbiography}

\begin{IEEEbiography}{Xinmiao Zhang}
received her Ph.D. degree in Electrical Engineering from the University of Minnesota. She joined The Ohio State University as an Associate Professor in 2017. Prior to that, she was a Timothy E. and Allison L. Schroeder Assistant Professor 2005-2010 and Associate Professor 2010-2013 at Case Western Reserve University. Between her academic positions, she was a Senior Technologist at Western Digital/SanDisk Corporation. Dr. Zhang's research spans the areas of VLSI architecture design, digital storage and communications, security, and signal processing.

Dr. Zhang received an NSF CAREER Award in January 2009. She is also the recipient of the Best Paper Award at 2004 ACM Great Lakes Symposium on VLSI and 2016 International SanDisk Technology Conference. She authored the book ``VLSI Architectures for Modern Error-Correcting Codes'' (CRC Press, 2015), and co-edited ``Wireless Security and Cryptography: Specifications and Implementations" (CRC Press, 2007). She also published more than 100 papers. She was elected to serve on the Board of Governers of the IEEE Circuits and Systems Society for the 2019-2021 term and is a member of the CASCOM and VSA technical committees. She was also a Co-Chair of the Data Storage Technical Committee (2017-2020). She served on the technical program and organization committees of many conferences, including ISCAS, SiPS, ICC, GLOBECOM, GlobalSIP, and GLSVLSI. She has been an associate editor for the IEEE Transactions on Circuits and Systems-I 2010-2019 and IEEE Open Journal of Circuits and Systems since 2019.
\end{IEEEbiography}

\end{document}